\documentclass[12pt]{iopart}
\eqnobysec
\usepackage{setstack}
\usepackage{iopams}
\usepackage{graphicx}

\begin{document}

\title[Fluctuations and correlations in biological coevolution]{Fluctuations 
and correlations in an individual-based model of biological coevolution}
\author{R K P Zia\dag\ddag \; and Per Arne Rikvold\P}
\address{\dag\ Center for Stochastic Processes in Science and Engineering, 
Department
of Physics, Virginia Polytechnic Institute and State University, Blacksburg,
Virginia 24061-0435, USA}
\address{\ddag\ Fachbereit Physik, Universit\"{a}t
Duisburg-Essen, 45117 Essen, Germany}
\address{\P\ School of Computational Science
and Information Technology, Center for Materials Research and Technology,
and Department of Physics, Florida State University, Tallahassee, Florida
32306-4120, USA}

\ead{rkpzia@vt.edu, rikvold@csit.fsu.edu}

\date{\today }

\begin{abstract}
We extend our study of a simple model of biological coevolution to its
statistical properties. Staring with a complete description in terms of a
master equation, we provide its relation to the deterministic evolution
equations used in previous investigations. The stationary states of the
mutationless model are generally well approximated by Gaussian
distributions, so that the fluctuations and correlations of the populations
can be computed analytically. Several specific cases are studied by Monte
Carlo simulations, and there is excellent agreement between the data and the
theoretical predictions.
\end{abstract}

\submitto{JPA}

\maketitle

\section{Introduction}
\label{sec:Int}

The dynamics of populations and species in the context of biological or
ecological systems have attracted considerable attention in the community of
statistical physicists in the last decade. Though there has been much
progress \cite{DROS01,LASS02}, it is rare that the behavior of macroscopic
populations can be predicted, even for ``simple models,'' from a set of 
{\em stochastic} rules for an {\em individual's} propensity to survive and/or
reproduce. The difficulties can be traced not only to the presence of many
degrees of freedom, with complex internal interactions (e.g., mutualistic or
predator-prey), but also to the non-trivial couplings to external reservoirs
(such as energy or food). As a result, even if we only focus on systems in
stationary states, we must recognize that these are {\em non-equilibrium
steady states}, so that the well-known methods of equilibrium thermodynamics
should not be blindly applied. At present, given the absence of a
universally applicable framework of non-equilibrium statistical mechanics,
progress is made through understanding ``one system at a time.'' Within this
context, we recently studied a model of coevolution \cite{RIKV03A,RIKZIA03},
based on the one (``tangled nature'') introduced by Hall, Christensen, and
collaborators \cite{HALL02,CHRI02,COLL03}. The motivations behind these
studies are varied. An early model for coevolution and speciation,
introduced by Bak and Sneppen \cite{BAK93}, consists of competing species,
according to a preassigned notion of ``fitness.'' Speciation arises by
imposing a crude version of ``mutation,'' i.e., letting the least fit
species (as well as some of their ``neighbors'') be replaced by new species
with different, randomly chosen fitness. Despite their simplicity, such models
appear to settle into a steady state which exhibits avalanches of
extinctions. Though they are hailed as showing a link between Darwin's
principle of ``survival of the fittest'' and Gould's notion of ``punctuated
equilibria" \cite{GOUL77,GOUL93,NEWM85}, these models are thought to be
too simplistic in at least two aspects. Firstly, mutation and selection act
on individuals of a populations, rather than on entire species at once.
Secondly, whether a species is ``fit'' is not a static notion, but rather
depends on what other species are present in the ecosystem.

To address both issues, Hall, {\em et.al.} \cite{HALL02,CHRI02,COLL03}
recently introduced an individual-based model with a dynamically evolving
``fitness landscape.'' The ``ecosystem'' consists of individuals, each 
of which is
said to belong to a ``species'' identified by a ``genome'' 
represented by a string of bits. Mutations are built in through the random
flipping of bits, at a constant slow rate, in the genomes of newborn 
individuals.
In earlier models of speciation, an individual's fitness, i.e., its 
reproduction
probability, is purely a function of the bit string and fixed for all time.
Here, the reproduction probability 
depends on the relative abundance of all other species,
so that ``fitness'' becomes a dynamic concept. The ``interspecies
interactions'' are, by contrast, comparatively static in nature and, for
simplicity, are introduced via fixed quantities associated with {\em pairs\/}
of genotypes. Thus, the model accounts only for whether (an individual of) a
species is beneficial or detrimental for another species, regardless of the
presence of any other species. Again for simplicity, all species reproduce
asexually, with identical fecundities. Simulations reveal several
interesting behaviours, including the presence of long-lived states separated
by bursts of high activity -- ``punctuated equilibria.'' Unfortunately, this
model proved too complex for analytic understanding, and only
phenomenological descriptions have been advanced to date.

In an effort to gain some insight into how its remarkable properties arise,
we considered a variation of this model, with simpler interspecies
interactions, which enables us both to carry out much longer simulations and
to perform linear stability analysis~\cite{RIKV03A,RIKZIA03}. In addition to
observing a self-similar picture of intermittency or ``punctuated
equilibria'' over several decades of generations, we found that standard
measures of diversity display $1/f$ noise in their power spectral densities.
This property is quite consistent with another finding: The life-time
distribution of the long-lived states approximately 
follows a inverse-square power law \cite{Krug03}.
More detailed investigations of the very long-lived states, which typically
consist of a community of a 
handful of dominant species along with a ``cloud of mutants,''
reveal the reason behind their longevity. Essentially none of the closest
mutants of the main species in such a community are ``dangerous,'' in that
their interactions with the parent species inhibit their exponential growth.
In other words, the original community is (linearly) stable against invasion 
by its most closely related mutants. 
Only a small fraction of the next-closest mutants (with
genes differing from a dominant species by two bits) are dangerous, leading
to the eventual demise of the ``quasi-steady'' state. Despite the
discovery of these connections, full analytic understanding is still beyond
our grasp. In particular, since the theoretical analyses were based entirely
on a heuristic, deterministic (``mean-field'') equation of motion, it is
unlikely that they can account for the most intriguing behaviour stemming
from a {\em stochastic} dynamics.

In the present paper, we address some of the issues associated with a fully
stochastic description. Starting from a master equation which governs the
evolution of all the details (``microscopics'') of the model, we first
demonstrate that the deterministic equation in \cite{RIKZIA03} emerges,
provided all correlations are ignored. However, it is no easy task to find a
quantitative understanding describing the quasi-steady states (QSS),
in which the system appears to be stationary but 
actually has long, finite lifetimes.
One complication lies with the inherent metastability aspect of a QSS.
Another is that, due to mutations, the populations in a QSS consist of 
{\em two} components: a handful of dominant species (with at least 
several hundred 
individuals, in the specific simulations we ran) and a larger number of
minor species (with much less than one hundred individuals). 
Our approach to the solution is
a two-step process. First, for each QSS, we develop a full understanding of
a corresponding ``truly stationary state'' (TSS), i.e., one with only the
dominant species. Each TSS is associated with an ${\cal N}$-species fixed
point discussed in \cite{RIKZIA03} and can be accessed by setting 
the mutation rate, $\mu $, to zero. The second step is to account for 
$O\left(\mu \right) $ effects, by including a limited ``cloud of mutants.'' 
Needless
to say, a careful definition of such a community will be necessary before
any analytic progress is possible. Thus, we will only take the first step
here, deferring the more complex problem to a later publication. Beyond
that, our eventual goal is to predict the more fascinating phenomena, such
as power-law distribution of QSS lifetimes, ``punctuated equilibria'' (or
intermittency), and $1/f$ noise.

In the next section, for completeness and the readers' convenience, we
briefly review the specifications of the model. The following section is
devoted to the master equation and the derivation of the deterministic
evolution equation used in \cite{RIKZIA03}. Fluctuations and
correlations in a TSS, as well as comparisons to simulations in specific
cases, are the focus of sections~\ref{sec:TSS} and~\ref{sec:dyn}.
Conclusions and an outlook can be found in section~\ref{sec:Concl}. The
Appendix is devoted to some of the technical details.

\section{Model specification and algorithm}

\label{sec:Mod}

The model we considered \cite{RIKZIA03} is a simplified version of the one
introduced in \cite{HALL02,CHRI02,COLL03}. It consists of a population of
individuals, each of which is associated with a string $L$ of bits (0 or 1),
representing a ``genome'' of $L$ ``genes.'' The bit-strings, or genotypes,
are labeled by integers $I$ , which lie between $1$ and 
${\cal N}_{{\rm max}}=2^L$. 
For simplicity, we will use the term ``species'' to distinguish
individuals with different bit-strings or genotypes. The population evolves
asexually in discrete time steps ($t=0,1,...$), which may be thought of as
``years'' or ``generations.'' In our model, all individuals of a generation
die when those of the next generation are ``born'' (as in, e.g., aphids).
Thus, for any particular run (or ``history''), the system is fully specified
by the set of integers $n_I(t)$ ($I=1,...,{\cal N}_{{\rm max}}$)
representing the number of individuals of genotype $I$ in generation $t$. In
cases where the explicit index $I$ is not necessary, we will use the
``vector'' notation: 
\begin{equation}
\vec{n}(t)\equiv \{n_1(t),...,n_{{\cal N}_{{\rm max}}}(t)\}\,\,.
\label{eq:n^arrow}
\end{equation}

To model competition and interspecies interactions, we let an individual die
with some non-vanishing probability {\em before} it reproduces, e.g., salmon
that die in the oceans. All survivors then give birth to $F$ offspring,
which constitute the next generation. Competition for resources (e.g.,
space, energy, food) is often introduced via a Verhulst \cite{VERH1838}
factor, which also plays the role of preventing unlimited growth and enters
typically via a ratio $N_{{\rm tot}}(t)/N_0$. Here, $N_0$ is a parameter
representing the ``carrying capacity'' of the ``ecosystem,'' and 
\begin{equation}
N_{{\rm tot}}(t)\equiv \sum_In_I(t)  \label{eq:N_totDef}
\end{equation}
is the total population at time $t$. With a ``healthy'' fecundity ($F$), 
the population is unlikely to ``collapse'' ($N_{{\rm tot}}=0$). Instead, 
$N_{{\rm tot}}(t)$ is rarely far from $N_0$. Interspecies interactions are
modeled by a matrix ${\bf M}$, the element $M_I^J$ being the effect of the
species $J$ on species $I$. For reasons provided in \cite{RIKZIA03}, we set 
$M_I^I=0$ and choose random off-diagonal elements from a uniform distribution
over $[-1,1]$. In all our simulations, ${\bf M}$ is fixed at $t=0$ and does 
{\em not} evolve in time. If both $M_I^J$ and $M_J^I$ are positive, the
two species are said to be ``mutualistic,'' and if they are of opposite
signs, we have a predator-prey relationship. Not surprisingly, populations
with both elements being negative are extremely unstable. Finally, the
probability of an individual of species $I$ to survive to reproduction is
specified by \cite{RIKZIA03} 
\begin{equation}
P(I;\vec{n}(t))=\frac 1{1+\exp \left[ -M_I^Jn_J(t)/N_{{\rm tot}}(t)+
N_{{\rm tot}}(t)/N_0\right] }\;,  
\label{eq:P}
\end{equation}
which is often shortened to just $P(I)$. Here, as in the rest of the paper,
we use the Einstein summation convention, i.e., a repeated index (e.g., $J$
in $M_I^Jn_J$) is summed over (from $1$ to ${\cal N}_{{\rm max}}$, in
general). As will be clear below, where we deal with systems with large 
$N_0$ (``macroscopic,'' though not necessarily in the sense of typical
thermodynamic systems), quantities with a single subscripted index (e.g., 
$n_J$) are generally of order $N_0$; those with a single superscript, of
order $1/N_0$; those with both (e.g., $M_I^J$) or none (e.g., $P(I)$), of 
$O\left( 1\right) $; etc. Exceptions are noted with a caret (or ``hat''). 
For example, we denote the ``normalized'' covariance matrix for the
populations (i.e., $\left( \left\langle n_In_J\right\rangle -\left\langle
n_I\right\rangle \left\langle n_J\right\rangle \right) /N_0$) by 
$\hat{G}_{IJ}$, which is a quantity of $O\left( 1\right) $ rather than 
$O\left(N_0^2\right) $. Similarly, its inverse ($\hat{\Gamma}^{IJ}$) is 
also of $O\left( 1\right) $ , as opposed to of $O\left( 1/N_0^2\right) $. 
To avoid confusion, we will remind the readers of such exceptions at the 
appropriate points.

The last ingredient in our model is mutation. In the absence of mutations,
the diversity of the population never increases with time. Indeed, the only 
{\em rigorously} stationary state is the collapsed one, $n_I = 0$ for all $I$. 
Nevertheless,
non-trivial states (with extremely long life times and called TSS's here)
are typically reached, for all {\em relevant} time scales. 
While the rigorously stationary state is independent of the intitial 
condition $\vec{n}(0)$, the TSS 
states are entirely dependent on the initial population, 
and they display the dominant characteristics of the corresponding QSS's.
Identified by a fixed point (FP) of the deterministic evolution equation, 
\begin{equation}
n_I(t+1)=n_I(t)FP(I;\vec{n}(t)\})\;,  \label{eq:EOM-no mu}
\end{equation}
a TSS provides the basis for linear stability analysis and for future
investigations of the associated QSS.

Returning to mutations, they are important, not only to promote diversity
and to model ``speciation,'' but also to provide the main ingredient for the
interesting phenomenon of, say, intermittency (``punctuated equilibria'') 
\cite{RIKZIA03}. As in all bit-string models involving asexual reproduction,
we allow each offspring to carry the genes of its parent, except for a
probability of $\mu /L$ that each bit be changed. As a result, on the
average and for small $\mu $, a survivor produces $\mu F$ offspring with
different genetic material. To keep track of the ``biodiversity,'' we define
the species richness ${\cal N}(t)$ as the number of populated species at $t$
(i.e., only ${\cal N}$ $I$'s are present at $t$). Another common measure
which characterizes the relative abundance of the species better (but will
not be the focus here) is the Shannon-Wiener index, $\sum_J\rho (J)\ln \rho
(J)$, where 
\begin{equation}
\rho (J)\equiv n_J(t)/N_{{\rm tot}}(t)  \label{eq:rhoDef}
\end{equation}
is just the fraction of species $J$ in the system.

Let us briefly summarize the algorithm used to simulate this model,
referring to \cite{RIKZIA03} for the details. There are three layers of
nested loops: (1) over generations $t$, (2) over ${\cal N}(t)$, and (3) over 
$n_I(t)$. In the innermost (last) loop, each individual produces, with
probability $P(I)$, $F$ offspring, each of which is allowed to mutate before 
$\vec{n}(t+1)$ is recorded. In most of our studies, we chose $L=13$, 
$N_0=2000$, $F=4$, $\mu =10^{-3}$, and $N_{{\rm tot}}\left( 0\right) =100$,
with random initial $\vec{n}(0)$. Thus, ${\cal N}_{{\rm max}}=8192$, though
far fewer species are typically present in the system at any given time,
i.e., ${\cal N}\ll {\cal N}_{{\rm max}}$.

\section{Master equation and mean-field theory}

\label{sec:MEMFT}

In our recent investigations, the Monte Carlo studies generate stochastic
sequences of configurations (i.e., non-negative integers, or just
``points,'' in the ${\cal N}_{{\rm max}}$-dimensional space) but the
theoretical analysis was based on deterministic, heuristic equations of
motion for the averages of the populations (``mean-field'' theory). To
understand the full stochastic process, we need a complete description,
involving the probability that the system is found with a specific number of
individuals at time $t$, namely, ${\cal P}\left( \vec{n},t\right) $. Its
evolution is governed by the master equation, given in an appendix of \cite
{RIKZIA03}. Before turning to this equation, let us emphasize the difference
between ${\cal P}\left( \vec{n},t\right) $ here and the $\vec{n}(t)$ above.
In the former, $\vec{n}$ is a co-ordinate (in ${\cal N}_{{\rm max}}$-space)
and ${\cal P}$ an evolving {\em function} in this space. By contrast, 
$\vec{n}(t)$ is just the trajectory of a single point in this space. A 
single Monte Carlo run generates a particular trajectory (or ``history''): 
${n}_I\left( t\right) $, and can be represented as a (Kronecker) delta 
function jumping from point to point: 
${\cal P}_{\mathrm{a\;MC\;run}}\left( \vec{n},t \right) = 
\prod_I\delta \left( n_I,{n}_I\left( t\right) \right)$.
The full dynamics of ${\cal P}\left( \vec{n},t\right) $ is simulated by
averaging over many runs, and thus difficult to access. When we turn out
attention to the TSS's below, the process simplifies, since they are
characterized by {\em static } distributions: 
${\cal P}^{*}\left( \vec{n}\right) $. Then, it is sufficient to perform a 
single, long run during which the system rarely wanders far (say, 
$O\left( \sqrt{N_0}\right) $) from the neighborhood of a specific point 
(typically, fixed points of the mean-field evolution equations).

Returning to the issue at hand, finding an equation for 
${\cal P}\left( \vec{n},t\right) $, 
we recapitulate the probability for an individual of species $I$ to survive: 
\begin{equation}
P\left( I\right) =\left\{ 1+\exp \left[ \frac{N_{{\rm tot}}}{N_0}
- M_I^J \frac{n_J}{N_{{\rm tot}}}\right] \right\} ^{-1} \; .  
\label{eq:B1}
\end{equation}
Again, note the different interpretation we give for this expression versus
equation
(\ref{eq:P}). Here, $P\left( I\right) =P\left( I;\vec{n}\right) $ denote 
${\cal N}_{{\rm max}}$ functions defined in ${\cal N}_{{\rm max}}$-space,
independent of $t$. In contrast, for a particular MC run at a particular
time $t$, we need only ${\cal N}\left( t\right) $ functions (for a specific
point $\{n_J(t)\}$, in a much smaller, ${\cal N}$-space). Next, we must keep
track of {\em all} of the possible number of survivors. Since all of these
individuals reproduce, we will call them ``parents.'' Defining the symbol 
${n_I \atopwithdelims[] m_I}$
as the rate for $m_I$ individuals to survive from
the original $n_I$, we simply write a binomial distribution: 
\begin{equation}
{n_I \atopwithdelims[] m_I} 
=\frac{n_I!}{m_I!\left( n_I-m_I\right) !}\left[
P\left( I\right) \right] ^{m_I}\left[ 1-P\left( I\right) \right] ^{n_I-m_I}.
\label{eq:B2}
\end{equation}
Next, each parent gives rise to $F$ offspring, not every one of which is of
the same species.  In the simulations, it is possible to have a mutant
whose genome differs from the parent by two or more bits. However, this is
quite rare, being less than 
$O\left( \mu ^2N_{{\rm tot}}\right) =O\left( \mu^2N_0\right) \sim 10^{-3}$. 
Thus, we will keep our analysis simple by
restricting our analysis here to mutations which flip only a single bit.
Then, there can be only $L+1$ possible varieties of offspring for each
parent. To account for these, let us introduce the notation 
\begin{eqnarray*}
&&b_{J,0}\quad \mathrm{for\;the\;number\;of\;offspring\;from\;parent}\;J\;
\mathrm{with\;no\;mutations} \\
&&b_{J,\alpha }\quad \mathrm{for\;the\;number\;of\;offspring\;from\;parent 
\;}J \;\mathrm{with\;the\;}\alpha \mathrm{th\;bit\;flipped}
\end{eqnarray*}
and define the multinomial-like symbol 
\begin{equation}
{Fm_J \atopwithdelims[] b_{J,0},b_{J,1},...,b_{J,L}} 
=\frac{\left(Fm_J\right) !}{\left( b_{J,0}\right) !}
\left( 1-\mu \right)^{b_{J,0}}
\prod_{\alpha =1}^L\frac 1{\left( b_{J,\alpha }\right) !}
\left(\frac \mu L\right) ^{b_{J,\alpha }} \; .  
\label{eq:B3}
\end{equation}
This is the probability that the $Fm_J$ offspring are distributed into the
specific set $\left\{ b_{J,0},b_{J,1},...,b_{J,L}\right\} $. The last
ingredient needed is the connection matrix 
\begin{equation}
\Delta _K^{J,\alpha }=\left\{ 
\begin{array}{cc}
1 & \mathrm{if\;genotype\;}K\;\mathrm{is\;}J\mathrm{\;with\;the\;}\alpha 
\mathrm{th\;bit\;flipped} \\ 
0 & \mathrm{otherwise}
\end{array}
\right.  \label{eq:B4}
\end{equation}
so that the number of offspring born {\em into} species $K$ due to 
mutations is 
\begin{equation}
B_K\equiv \sum_{J,\alpha >0}\Delta _K^{J,\alpha }b_{J,\alpha }\;.
\label{eq:B5}
\end{equation}
With these ingredients, we arrive at the master equation
\footnote{
Many master equations are written for continuous time, in the form of 
$\partial _t{\cal P}(C,t)=\sum_{C^{\prime }}L\left( C,C^{\prime }\right) 
{\cal P}(C^{\prime },t)$. For discrete-time processes, we find it more
convenient to express the evolution in the form used here.}: 
\begin{equation}
\fl
{\cal P}\left( \vec{n}^{\prime },t+1\right) 
=\sum_{ \vec{n},\vec{m},\left\{b\right\} }
\prod_K 
\delta \left( n_K^{\prime },b_{K,0}+B_K \right) 
\prod_J
{Fm_J \atopwithdelims[] b_{J,0},b_{J,1},...,b_{J,L}} 
\prod_I 
{n_I \atopwithdelims[] m_I} 
{\cal P}\left( \vec{n},t \right) \;,  
\label{eq:B6}
\end{equation}
where $\delta \left( n^{\prime },n\right) $ is the Kronecker delta. 
Given a particular initial configuration $\vec{n}_0$, 
${\cal P}\left( \vec{n},t\right) $ can be found, in principle, by recursion 
with ${\cal P}\left(\vec{n},0\right)=\delta \left( \vec{n},\vec{n}_0\right)$. 
In that sense, we note that the more precise notation is 
${\cal P}\left( \vec{n},t|\vec{n}_0,0\right) $, 
explicitly showing that it is the probability to find our
system in state $\vec{n}$ at time $t$, {\em conditioned} on a specific
initial condition. However, this notation seems unnecessarily cumbersome, so
that we will just use ${\cal P}\left( \vec{n},t\right) $ in its place.

Note that equation (\ref{eq:B6}) is just a special example of the general
evolution of conditional probabilities in a Markov process, i.e., 
\begin{equation}
{\cal P}\left( \vec{n}^{\prime },t+1\right) =\sum_{\vec{n}}R\left( 
\vec{n}^{\prime }|\vec{n}\right) {\cal P}\left( \vec{n},t\right) \;,
\label{eq:MEgen}
\end{equation}
where $R\left( \vec{n}^{\prime }|\vec{n}\right) $ is the conditional
probability for finding the system in $\vec{n}^{\prime }$ given that it was
in $\vec{n}$, also known as the transition rate. For our case, $R$ is
explicitly 
\begin{equation}
R\left( \vec{n}^{\prime }|\vec{n}\right) =\sum_{\vec{m},\left\{ b\right\}}
\prod_K \delta \left( n_K' , b_K+B_K \right) \prod_J  
{Fm_J \atopwithdelims[] b_{J,0},b_{J,1},...,b_{J,L}} 
\prod_I 
{n_I \atopwithdelims[] m_I} 
\;.
\label{eq:R}
\end{equation}

Once ${\cal P}\left( \vec{n},t\right) $ is found, the time dependence of the
expectation value of any quantity can be obtained via 
\begin{equation}
\left\langle \bullet \right\rangle _t\equiv \sum_{\vec{n}}\bullet {\cal P}
\left( \vec{n},t\right) \,\,,  
\label{eq:B7}
\end{equation}
in principle. Specifically, our main interest here will be $\left\langle
n_I\right\rangle _t$, the average number of individuals of species $I$ at
time $t$, as well as quantities quadratic in $n$ (e.g., covariances 
and correlations). For
example, to see how $\left\langle n_K\right\rangle _t$ evolves in time, we
multiply equation (\ref{eq:MEgen}) by $n_K^{\prime }$, and sum over 
$\vec{n}^{\prime }$
\begin{equation}
\left\langle n_K\right\rangle _{t+1}=\left\langle f_K\left( \vec{n}\right)
\right\rangle _t\,\,,  
\label{eq:nEvolve}
\end{equation}
where 
\begin{equation}
f_K\left( \vec{n}\right) \equiv \sum_{\vec{n}^{\prime }}n_K^{\prime }R\left( 
\vec{n}^{\prime }|\vec{n}\right) \,\,.  
\label{eq:fDef}
\end{equation}
Exploiting the explicit form of $R$ above and 
\begin{equation}
\sum_mm\frac{n!}{m!\left( n-m\right) !}q^m\left( 1-q\right) ^{n-m}=mq\,\,,
\end{equation}
it is straightforward to find $f$. Reminding the readers of the $\vec{n}$
dependence in $P$, we write explicitly 
\begin{equation}
f_K\left( \vec{n}\right) =F\left[ \left( 1-\mu \right) n_KP\left( K;\vec{n}
\right) +\left( \mu /L\right) \sum_J\sum_{\alpha =1}^L\Delta _K^{J,\alpha
}n_JP\left( J;\vec{n}\right) \right] \,\,.  \label{eq:f}
\end{equation}
The interpretation of various terms in this expression is clear: $F$ offspring
are born to $n_JP\left( J\right) $ survivors of species $J$, with
rearrangements into the new generations due to mutations. Inserting it into
equation (\ref{eq:nEvolve}), we arrive at an {\em exact} equation 
\begin{equation}
\left\langle n_K\right\rangle _{t+1}=F\left[ \left( 1-\mu \right)
\left\langle n_KP\left( K\right) \right\rangle _t+\left( \mu /L\right)
\sum_J\sum_{\alpha =1}^L\Delta _K^{J,\alpha }\left\langle n_JP\left(
J\right) \right\rangle _t\right] \,\,.  \label{eq:nExact}
\end{equation}
Its simplicity is deceptive, since the $P$'s are, from equation (\ref{eq:B1}),
non-trivial functions of $\vec{n}$. If a formal expansion in powers of 
$\vec{n}$ were inserted for these functions, then averages of all powers, 
$\left\langle n_Jn_K...\right\rangle _t$, will appear on the right-hand side.
Of course, equations for these new averages can be written formally, but the
result would be more complex than the BBGKY hierarchy \cite{BBGKY}. Needless
to say, good approximation schemes are crucial for further progress.

One such scheme, also known as the ``mean-field'' approximation, is to
ignore all correlations in order to produce a closed equation for 
$\left\langle n_K\right\rangle $. Thus, we replace all averages of the
products by products of the averages: 
\begin{equation}
\left\langle n_J n_K... \right\rangle \rightarrow \left\langle
n_J\right\rangle \left\langle n_K\right\rangle ... 
\end{equation}
so that, e.g., 
\begin{equation}
\left\langle n_KP \left( K ; \vec{n} \right) \right\rangle \rightarrow
\left\langle n_K\right\rangle P\left( K;\left\langle \vec{n}\right\rangle
\right) \;. 
\end{equation}
In terms of the less cumbersome notation $n_K\left( t\right) \equiv
\left\langle n_K\right\rangle _t$, equation (\ref{eq:nEvolve}) reduces to 
\begin{equation}
n_K(t+1)=f_K\left( \{n_J\left( t\right) \}\right) \,\,.  
\label{eq:MFT}
\end{equation}
Apart from the change of notation ($P_K(\{n_J(t)\})$ 
$\rightarrow P(K;\vec{n}(t))$), 
this equation is precisely the starting point of our earlier
analysis, equation~(2) in \cite{RIKZIA03}: 
\begin{equation}
\fl
n_K(t+1)=n_K(t)FP_K(\{n_J(t)\})[1-\mu ]
+
(\mu/L)F\sum_{I(K)}n_{I(K)}(t)P_{I(K)}(\{n_J(t)\})\;,  
\label{eq:EOM}
\end{equation}
where $I\left( K\right) $ runs over all values of $I$ that differ from $K$
by one bit.

In the low-mutation-rate ($\mu \ll 1$) regime, much of the behaviour of a
QSS is well approximated by a TSS ($\mu =0$), which can be understood in
terms of the fixed points of the mutationless version of this mean-field
theory. Here, let us briefly summarize the predictions of this theory. With 
$\mu =0$, the number of populated species in the system, ${\cal N}$, never
increases. Therefore, we can restrict our attention to a small subspace 
(${\cal N}$-dimensional, with ${\cal N}<5$ in typical simulation runs) of 
the full $2^L$-dimensional space. In our previous work \cite{RIKZIA03}, we 
use tildes (e.g., ${\bf \tilde{M}}$) to emphasize this aspect. To keep the
notation simple, we will drop the tildes here and keep in mind that, for
example, ${\bf M}$ is an ${\cal N}\times {\cal N}$ matrix. Also, with 
${\cal N}$ being an $O\left( 1\right) $ quantity (i.e., small compared to 
$N_0$), the population of every species is generically $O\left( N_0\right)$.

Next, $f_I$ simplifies to 
\begin{equation}
f_I\left( \vec{n}\right)  
{\underset {\mu =0} \rightarrow } 
Fn_IP\left( I;\vec{%
n}\right) \,\,.  
\label{eq:f(n)}
\end{equation}
Since the fixed point, $\vec{n}^{*}$, obeys 
\begin{equation}
\vec{n}^{*}=\vec{f}\left( \vec{n}^{*}\right) \,\,,  
\label{eq:n*eqn}
\end{equation}
we have $FP\left( I;\vec{n}^{*}\right) =1$ for all $I$. Defining the inverse
of ${\bf M}$ by 
\begin{equation}
{\bf W}\equiv {\bf M}^{-1}\,\,,  
\label{eq:W}
\end{equation}
(with elements $W_I^J$) and the sum 
\begin{equation}
\sigma \equiv \sum_{IJ}W_I^J\,\,,  \label{eq:Sigma}
\end{equation}
we found that the fractions of each species are given by 
\begin{equation}
\rho ^{*}\left( I\right) \equiv n_I^{*}/N_{{\rm tot}}^{*}=\sum_JW_I^J/\sigma
\,\,,  \label{eq:fix3}
\end{equation}
and the total $N_{{\rm tot}}^{*}$, by 
\begin{equation}
N_{{\rm tot}}^{*}/N_0=\ln (F-1)+1/\sigma \;.  \label{eq:fix4}
\end{equation}
Finally, the elements of ${\bf S}$, the stability matrix, is 
\begin{equation}
S_I^J\equiv f_I{}^J\left( \vec{n}^{*}\right) \equiv \left. \partial
f_I/\partial n_J\right| _{\vec{n}^{*}}=\delta _I^J+\Lambda _I^J\,\,,
\label{eq:S}
\end{equation}
where $\delta _I^J$ is the unit matrix. Here, 
\begin{equation}
\Lambda _I^J=\left( 1-\frac 1F\right) \left( M_I^J-\ln (F-1)-2/\sigma
\right) \rho ^{*}\left( I\right) \;.  
\label{eq:A2}
\end{equation}
(denoted by $\tilde{\Lambda}_{IJ}$ in \cite{RIKZIA03} and referred to as the
community matrix in biological literature) plays the role of a ``restoring''
force, driving the population back towards the (stable) fixed point.

\section{Fluctuations and correlations of populations}
\label{sec:TSS}

It is clear that all stochastic aspects of the system are lost in the
mean-field approximation. In this section, we study the simplest aspect,
namely, an approximate description of the long-lived, quasi-steady
states (QSS). Now, as we have seen in simulations \cite{RIKZIA03}, these
states are dominated by a few mutually supportive species, along with a few
individuals of closely related ``benign mutants.'' Thus, we restrict
ourselves in this paper to systems {\em without mutation}, so that several
simplifications apply. First, there is the obvious reduction of $R$ from
equation (\ref{eq:R}) to 
\begin{equation}
R \left( \vec{n}^{\prime } | \vec{n} \right) 
{\underset {\mu =0} \rightarrow } 
\sum_{\vec{m}} \prod_K \delta \left( n_K^{\prime },Fm_K\right) 
\prod_I 
{n_I \atopwithdelims[] m_I} 
\,\,.
\label{eq:R0}
\end{equation}
Next, since there can be no new species, we can focus on populations with 
${\cal N}\sim O\left( 1\right) $ species, each of which having $O\left(
N_0\right) $ individuals. Finally, though it is possible for a fluctuation
to collapse the entire population (which is, rigorously, {\em the unique}
stationary state associated with equations~(\ref{eq:MEgen},\ref{eq:R0})), such
events are so extreme that the life times of a typical non-trivial state
should be $O\left( \rme^{N_0}\right) $. So, for all practical purposes, we may
regard such states as ``truly'' stationary (the TSS's).
\footnote{
It is possible to force such states to be rigorously stationary, by slight
modifications of the rates. The simplest is to let the survival
probabilities $P\left( I\right) $ be unity when $n_I=1$.}

A complete description of a TSS is provided by a stationary 
{\em distribution} ${\cal P}^{*}\left( \vec{n}\right) $, which satisfies 
\begin{equation}
{\cal P}^{*}\left( \vec{n}^{\prime }\right) =\sum_{\vec{n}}R\left( \vec{n}
^{\prime }|\vec{n}\right) {\cal P}^{*}\left( \vec{n}\right) \,\,.
\label{eq:P*}
\end{equation}
Given ${\cal P}^{*}\left( \vec{n}\right) $, we can find stationary averages
of any quantity: 
\begin{equation}
\left\langle \bullet \right\rangle ^{*}\equiv \sum_{\vec{n}}\left( \bullet
\right) {\cal P}^{*}\left( \vec{n}\right) \,\,.  \label{eq:<dot>}
\end{equation}
Here, our main interest will be mean populations and their correlations: 
\begin{equation}
\left\langle n_I\right\rangle ^{*}\quad \mathrm{and}\quad \left\langle
n_In_J\right\rangle ^{*}-\left\langle n_I\right\rangle ^{*}\left\langle
n_J\right\rangle ^{*}\,\,.  
\label{eq:n+C}
\end{equation}

Even in the absence of mutations and with simplified $R$'s, there are
substantial non-linearities (through $P\left( \vec{n}\right) $) in the
problem that prevent us from finding solutions to equation (\ref{eq:P*}) in
general. On the other hand, observations in simulations (and lessons learned
from the central limit theorem) show that our distributions are well
approximated by Gaussians. Indeed, a systematic expansion for 
${\cal P}^{*}$ can be formulated, starting from the Gaussian form: 
\begin{equation}
\fl
{\cal P}_{\rm G} \left( \vec{n}\right) 
=
\left( 2\pi N_0\right) ^{-{\cal N}/2}\left(
\det {\bf \hat{\Gamma}}\right) ^{1/2}\exp \left[ -\frac 1{2N_0}\left( n_I-
\bar{n}_I\right) \hat{\Gamma}^{IJ}\left( n_J-\bar{n}_J\right) \right] \,\,.
\label{eq:PG}
\end{equation}
where $\bar{n}_I$ and $\hat{\Gamma}^{IJ}$ are parameters to be determined.
Of course, the first of these is of $O\left( N_0\right) $. As for the
latter, we expect the fluctuations in our problem to be 
$O\left( \sqrt{N_0}\right) $, i.e., covariances of $O\left( N_0\right) $. 
To incorporate this
expectation, we have chosen to write ${\cal P}_{\rm G}$ in a form such that 
the matrix ${\bf \hat{\Gamma}}$ has elements of $O\left( 1\right) $: 
\begin{equation}
\hat{\Gamma}^{IJ}\sim O\left( 1\right) \,\,,  \label{eq:Gam}
\end{equation}
{\em despite} the presence of two superscripts.

Within the context of such a scheme, it is possible to compute these
quantities from the microscopic rates. Before proceeding to this
computation, we remark that this approximation for ${\cal P}^{*}$ will lead
to the predictions 
\begin{equation}
\left\langle n_I\right\rangle ^{*}=\bar{n}_I\quad \mathrm{and}\quad
\left\langle n_In_J\right\rangle ^{*}-\left\langle n_I\right\rangle
^{*}\left\langle n_J\right\rangle ^{*}=N_0\hat{G}_{IJ}\,,  
\label{eq:nAv+cov}
\end{equation}
where ${\bf \hat{G}}$ is the inverse of ${\bf \hat{\Gamma}}$, i.e., 
\begin{equation}
\quad \hat{G}_{IJ}\hat{\Gamma}^{JK}=\delta _I^K\,\,.  
\label{eq:GGam=1}
\end{equation}
Of course, we expect 
\begin{equation}
\hat{G}_{IJ}\sim O\left( 1\right) \,\,,  
\label{eq:G}
\end{equation}
despite its two subscripts.

Though such approaches are well known \cite{vanKampen}, we provide a few
details in the Appendix, both for the sake of completeness and for the
convenience of readers unfamiliar with these methods. Here, we present only
the highlights of the analysis and how they apply to our case. As shown
below, the agreement between our predictions and simulation data in 
three specific
cases are excellent and validates this entire approach.

Turning to the computation of $\bar{n}_I$ and ${\bf \hat{\Gamma}}$ (or 
${\bf \hat{G}}$), we may expect the former to be simply related to the 
$n_I^{*}$ of mean-field theory (equation (\ref{eq:n*eqn})). As shown in the 
appendix, we have 
\begin{equation}
\bar{n}_I=n_I^{*}\left[ 1+O\left( 1/N_0\right) \right]  
\label{eq:nbar=n*}
\end{equation}
(provided none of the eigenvalues of ${\bf S}$ are close to unity). Indeed,
the first correction can be computed explicitly. Since these corrections are
not necessary for finding the fluctuations and correlations, we will not
quote the results here, but refer the reader to equation (\ref{eq:n-G}) in the
Appendix. To find ${\bf \hat{G}}$, we need not only 
$f_I\left( \vec{n}\right) $ and the stability matrix ${\bf S}$ 
(given by equations (\ref{eq:f(n)},\ref{eq:S},\ref{eq:A2})), but also 
\begin{equation}
H_{IJ}\left( \vec{n}\right) \equiv \sum_{\vec{n}^{\prime }}n_I^{\prime
}n_J^{\prime }R\left( \vec{n}^{\prime }|\vec{n}\right) \,\,.
\end{equation}
In the limit $\mu \rightarrow 0$, $R$ reduces to equation (\ref{eq:R0}), 
so that 
\begin{eqnarray}
H_{IJ}\left( \vec{n}\right) &\rightarrow &\sum_{\vec{m}}\left( Fm_I\right)
\left( Fm_J\right) \prod_I
{n_I \atopwithdelims[] m_I} 
\\
&=&F^2\left[ n_In_JP\left( I\right) P\left( J\right) 
+ \delta_I^J n_IP\left( I\right) \left( 1-P\left( I\right) \right) \right]
\,\,,  
\label{eq:H(n)}
\end{eqnarray}
which is indeed of the form in equation (\ref{eq:h_IJ}): 
$H_{IJ}=f_If_J+N_0\hat{H}_{IJ}$. Thus, we readily identify $\hat{H}_{IJ}$ 
of equation (\ref{eq:Htilde}): 
\begin{equation}
\hat{H}_{IJ}=\delta_I^J F^2n_I^{*}P\left( I;\vec{n}^{*}\right)
\left[ 1-P\left( I;\vec{n}^{*}\right) \right] /N_0\,\,.
\label{eq:Htilde-here}
\end{equation}
But, $FP\left( I;\vec{n}^{*}\right) =1$ for all $I$, so that 
\begin{equation}
\hat{H}_{IJ}=\delta_I^J \left( F-1\right) n_I^{*}/N_0\,\,.
\end{equation}
Referring to the Appendix again for the details, ${\bf \hat{G}}$ can be
obtained from ${\bf S}$ and ${\bf \hat{H}}$ via the linear relationship 
\begin{equation}
{\bf \hat{G}-S\hat{G}S}^T={\bf \hat{H}\,\,.}  \label{eq:GSH}
\end{equation}
Inserting the explicit form for ${\bf \hat{H}}$ into 
equations (\ref{eq:Hab}, \ref{eq:explicit}), 
we arrive at a complete solution for the covariance matrix
in terms of $\lambda _a,u_a^K$, and $v_I^a$ (respectively, the eigenvalues
and the left and right eigenvectors of $S_K^I$, normalized by $%
u_a^Iv_I^b=\delta _a^b$) : 
\begin{equation}
\hat{G}_{IJ}=\left( F-1\right) \sum_{a,b,K}v_I^av_J^bu_a^Ku_b^K
\frac{n_K^{*}/N_0}{1-\lambda _a\lambda _b}+O\left( 1/N_0\right) \,\,.
\label{eq:explicitG}
\end{equation}

\begin{figure}[t]
\includegraphics[angle=0,width=.47\textwidth]{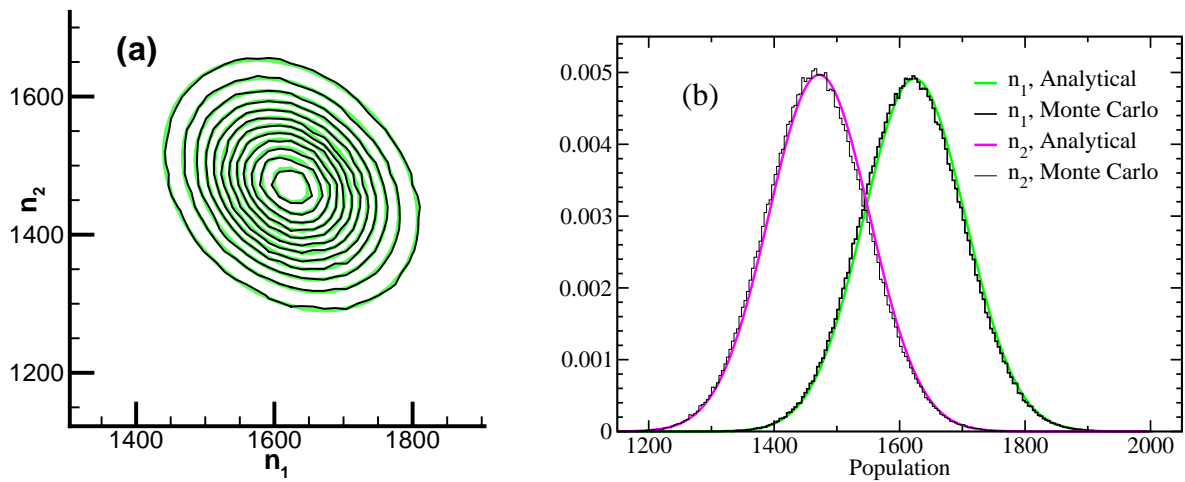}
\includegraphics[angle=0,width=.53\textwidth]{pops_hist.eps}
\caption[]{
Theoretical Gaussian (thick, grey curves in the background) and simulated 
(thin, black curves in the foreground) 
TSS probability densities for ${\cal N} = 2$, $F=4$, and
$N_0 = 2000$, corresponding to the second column in table~\ref{tab:I}. 
The Monte Carlo
simulation was performed over 524,290 generations with zero probability
of mutations. 
({\it a})
Contour plot of the 
joint distribution for $n_1$ and $n_2$, 
equation (\protect\ref{eq:PG}). 
The negative slope of the long
axis indicates that $n_1$ and $n_2$ are negatively correlated. 
({\it b})
Marginal distributions for $n_1$ (right) and $n_2$ (left),
equation (\protect\ref{eq:Pproj}). 
}
\label{fig:pops}
\end{figure}

Before comparing these predictions to simulations, we provide an intuitive
picture for these fluctuations and correlations. Since we are concerned with
distributions well approximated by Gaussians, the underlying process is just
an Ornstein-Uhlenbeck one \cite{vanKampen}. 
Such a process can be described by a Langevin
equation: 
\begin{equation}
\vec{\varepsilon}\left( t+1\right) -\vec{\varepsilon}\left( t\right) =
{\bf \Lambda }\vec{\varepsilon}\left( t\right) +\vec{\eta}\left( t\right) \;,
\label{eq:Langevin}
\end{equation}
where ${\bf \Lambda }$ plays the role of restoring forces which drive 
$\vec{\varepsilon}$ towards zero, and $\vec{\eta}$ is a Gaussian noise with 
\begin{equation}
\left\langle \vec{\eta}\left( t\right) \right\rangle =0;\,\,\left\langle 
\vec{\eta}\left( t\right) \vec{\eta}\left( t^{\prime }\right) \right\rangle 
= {\bf \hat{H}}\delta \left( t,t^{\prime }\right) \,\,.
\end{equation}
The notation for the two matrices (${\bf \Lambda }$, ${\bf \hat{H}}$) is
chosen deliberately. For this problem, they are precisely those given above:
equations (\ref{eq:A2},\ref{eq:Htilde-here}), while 
$\vec{\varepsilon}\left( t\right) $ is just the deviation from the average: 
\begin{equation}
\vec{\varepsilon}\left( t\right) =\vec{n}\left( t\right) -\bar{n}\,\,.
\label{eq:epsilon}
\end{equation}
The deterministic part of equation (\ref{eq:Langevin}) is intuitively clear,
being the same as in the mean-field evolution equation, linearized about the
fixed point. Now, $\vec{\eta}$ can be seen as the (microscopic) noise in the
stochastic process. In our case, this is clearly due to the uncertainties
associated with survival. Since we have a simple two-state (dying or
surviving) random process, we can hardly be surprised by the presence of the
factors $nP\left( 1-P\right) $. Also, since this randomness is imposed on
each individual, the diagonal form of ${\bf \hat{H}}$ can be anticipated.
Perhaps only the factor $F^2$ cannot be easily surmised.

\Table{
\label{tab:I}
Theoretical results
for two TSS communities with ${\cal N} = 2$ and~3, respectively, compared
with corresponding quantities from Monte Carlo simulations with $N_0 = 2000$ 
over 524,290 generations. The communities are defined by the interaction
matrices $\bf M$. 
The quantities shown are the normalized fixed-point population vector, 
$\vec{n}^*/N_0$ [equations (\ref{eq:fix3}) and~(\ref{eq:fix4})] 
and the corresponding Monte Carlo average 
$\langle \vec{n} \rangle_{\rm MC} /N_0$, 
the normalized population covariance matrix $\bf \hat{G}$ 
[defined by equation (\ref{eq:nAv+cov}) and calculated from equation
(\protect\ref{eq:recrel}) with 20 iterations for ${\cal N} =2$ and 30
iterations for ${\cal N} =3$] 
and ${\bf \hat{G}}_{\rm MC}$,
the step-covariance matrix $\bf \hat{g} $ [equation (\ref{eq:g})] and 
${\bf \hat{g}}_{\rm MC}$,
and the normalized 
correlation matrix between steps $\vec{s}(t)$ and the deviation from the 
average population $\vec{\varepsilon}(t) = \vec{n}(t) - \vec{n}^*$, 
$\bf \hat{C}$ [equation (\ref{eq:se})] and ${\bf \hat{C}}_{\rm MC}$. 
All numbers are given to four significant digits. 
Results for ${\cal N} = 4$ are shown in table~\protect\ref{tab:II}. 
}
\\
\br
\begin{tabular}{ccc}
${\cal N}$ & 2 & 3 
\\
\mr
\ms
$\bf M$ 
& 
$\left(
\begin{array}{cc}
0 & 0.9448 \\
0.8563 & 0
\end{array}
\right) $
& 
$\left(
\begin{array}{ccc}
0      & 0.7497 & 0.9450 \\
0.8935 & 0      & 0.6212 \\
0.6474 & 0.9881 & 0
\end{array}
\right) $
\\
\ms
\hline 
\ms
$\vec{n}^*/N_0$ 
& 
$\left(
\begin{array}{c}
0.8119 \\
0.7359
\end{array}
\right) $
& 
$\left(
\begin{array}{c}
0.6062 \\
0.4897 \\
0.5388
\end{array}
\right) $
\\
\ms
\hline
\ms
$\langle \vec{n} \rangle_{\rm MC} /N_0$ 
& 
$\left(
\begin{array}{c}
0.8112 \\
0.7350
\end{array}
\right) $
& 
$\left(
\begin{array}{c}
0.6055 \\
0.4892 \\
0.5378
\end{array}
\right) $
\\
\ms
\hline
\ms
$\bf \hat{G}$ 
& 
$\left(
\begin{array}{cc}
3.294& -0.8722 \\
-0.8722 & 3.224
\end{array}
\right) $
& 
$\left(
\begin{array}{ccc}
3.667   & -0.9323 & -0.9558 \\
-0.9323 & 3.551   & -0.9249 \\
-0.9558 & -0.9249 & 3.590
\end{array}
\right) $
\\
\ms
\hline
\ms
${\bf \hat{G}}_{\rm MC}$ 
& 
$\left(
\begin{array}{cc}
3.295& -0.8784 \\
-0.8784 & 3.227
\end{array}
\right) $
& 
$\left(
\begin{array}{ccc}
3.690   & -0.9440 & -0.9584 \\
-0.9440 & 3.573   & -0.9289 \\
-0.9584 & -0.9289 & 3.592
\end{array}
\right) $
\\
\ms
\hline
\ms
$\bf \hat{g} $ 
& 
$\left(
\begin{array}{cc}
4.455 & 1.368 \\
1.368 & 3.882
\end{array}
\right) $
& 
$\left(
\begin{array}{ccc}
3.039 & 0.7902 & 0.8768 \\
0.7902 & 2.285 & 0.7152 \\
0.8768 & 0.7152 & 2.592
\end{array}
\right) $
\\
\ms
\hline
\ms
${\bf \hat{g}}_{\rm MC}$ 
& 
$\left(
\begin{array}{cc}
4.442 & 1.352 \\
1.352 & 3.865
\end{array}
\right) $
& 
$\left(
\begin{array}{ccc}
3.050 & 0.7865 & 0.8716 \\
0.7865 & 2.287 & 0.7098 \\
0.8716 & 0.7098 & 2.586
\end{array}
\right) $
\\
\ms
\hline
\ms
$\bf \hat{C}$ 
& 
$\left(
\begin{array}{cc}
-2.227 & -0.6494 \\
-0.7189 & -1.941
\end{array}
\right) $
& 
$\left(
\begin{array}{ccc}
-1.520  & -0.5253 & -0.2812\\
-0.2649 & -1.143  & -0.5244\\
-0.5956 & -0.1908 & -1.296
\end{array}
\right) $
\\
\ms
\hline
\ms
${\bf \hat{C}}_{\rm MC}$ 
& 
$\left(
\begin{array}{cc}
-2.221 & -0.6385 \\
-0.7139 & -1.933
\end{array}
\right) $
& 
$\left(
\begin{array}{ccc}
-1.524  & -0.5221 & -0.2814\\
-0.2643 & -1.143  & -0.5224\\
-0.5902 & -0.1873 & -1.293
\end{array}
\right) $
\\
\ms
\br
\end{tabular}
\endTable

\Table{
\label{tab:II}
Theoretical results
for a TSS community with ${\cal N} = 4$, compared
with corresponding quantities from a Monte Carlo simulation with 
$N_0 = 10,000$ over 524,290 generations. 
The quantities shown are the same as in table~\protect\ref{tab:I}. 
The eigenvalues of $\bf S$ are relatively close to unity,
so evaluation of $\bf G$ from equation (\protect\ref{eq:recrel}) required
180 iterations. 
}
\\
\br
\ms
\begin{tabular}{cc}
${\cal N}$ & 4 
\\
\ms
\mr
\ms
$\bf M$ 
& 
$\left(
\begin{array}{cccc}
0 & 0.5507 & 0.5101 & 0.9847 \\
0.9543 & 0 & 0.8437 & 0.9508 \\
0.4724 & 0.6190 & 0 & 0.7371 \\
0.9734 & 0.6808 & 0.1862 & 0
\end{array}
\right) $
\\
\ms
\hline 
\ms
$\vec{n}^*/N_0$ 
& 
$\left(
\begin{array}{c}
0.3355 \\
0.7034 \\
0.2366 \\
0.3468
\end{array}
\right) $
\\
\ms
\hline
\ms
$\langle \vec{n} \rangle_{\rm MC} /N_0$ 
& 
$\left(
\begin{array}{c}
0.3356 \\
0.7035 \\
0.2354 \\
0.3472
\end{array}
\right) $
\\
\ms
\hline
\ms
$\bf \hat{G}$ 
& 
$\left(
\begin{array}{cccc}
4.215  & -1.043 & -2.583 & 1.132 \\
-1.043 & 3.882 & -0.1463 & -1.006 \\
-2.583 & -0.1463 & 6.509 & -3.791 \\
1.132 & -1.006 & -3.791 & 5.653 
\end{array}
\right) $
\\
\ms
\hline
\ms
${\bf \hat{G}}_{\rm MC}$ 
& 
$\left(
\begin{array}{cccc}
4.252  & -1.055 & -2.620 & 1.152 \\
-1.055 & 3.870 & -0.1366 & -1.003 \\
-2.620 & -0.1366 & 6.612 & -3.879 \\
1.151 & -1.003 & -3.879 & 5.718 
\end{array}
\right) $
\\
\ms
\hline
\ms
$\bf \hat{g} $ 
& 
$\left(
\begin{array}{cccc}
1.387 & 0.6468 & 0.2384 & 0.3094 \\
0.6468 & 3.704 & 0.4482 & 0.6518 \\
0.2384 & 0.4482 & 0.8927 & 0.2414 \\
0.3094 & 0.6518& 0.2414 & 1.461
\end{array}
\right) $
\\
\ms
\hline
\ms
${\bf \hat{g}}_{\rm MC}$ 
& 
$\left(
\begin{array}{cccc}
1.388 & 0.6417 & 0.2368 & 0.3090 \\
0.6417 & 3.694 & 0.4440 & 0.6501 \\
0.2368 & 0.4440 & 0.8870 & 0.2381 \\
0.3090 & 0.6501& 0.2381 & 1.462
\end{array}
\right) $
\\
\ms
\hline
\ms
$\bf \hat{C}$ 
& 
$\left(
\begin{array}{cccc}
-0.6934 & -0.3951 & -0.07285 & -0.1841 \\
-0.2517 & -1.852 & -0.1801 & -0.3281 \\
-0.1655 & -0.2681 & -0.4464 & -0.02048 \\
-0.1253 & -0.3237 & -0.2209 & -0.7304
\end{array}
\right) $
\\
\ms
\hline
\ms
${\bf \hat{C}}_{\rm MC}$ 
& 
$\left(
\begin{array}{cccc}
-0.6942 & -0.3907 & -0.07252 & -0.1851 \\
-0.2510 & -1.847 & -0.1779 & -0.3259 \\
-0.1642 & -0.2661 & -0.4435 & -0.01783 \\
-0.1239 & -0.3242 & -0.2202 & -0.7310
\end{array}
\right) $
\\
\ms
\br
\end{tabular}
\endTable

We have carried out simulations for ten cases, with ${\cal N}=2,3$, and 
$4$. For simplicity, in each case we chose a set of species which served 
as the dominant ones in one of the ten QSS communities 
included in Table~I of \cite{RIKZIA03}. Of
course, since $\mu $ was set to zero in the simulations reported here, 
the communities are TSS's.
The runs were carried out for 524,290 generations each. From the recorded 
$n_I\left( t\right) $, we computed the averages 
$\left\langle n_I\right\rangle_{\rm MC}$ and the covariance matrix 
${\bf \hat{G}}_{\rm MC}$. For the ${\cal N}=2$ case, we can easily display a 
histogram of all the populations
(figure~\ref{fig:pops}(a)). 
In addition, in figure~\ref{fig:pops}(b), we show how good the Gaussian
approximation is by plotting the projections of both this histogram (onto
one or the other axis) and the theoretical predictions, e.g., 
\begin{eqnarray}
{\cal P}_{\rm proj}\left( n_1\right) &=&\int \rmd n_2{\cal P}_{\rm G}
\left( n_1,n_2\right)
\\
&=&\left( \frac{\det {\bf \hat{\Gamma}}}{2\pi N_0\hat{\Gamma}_{22}}\right)
^{1/2}\exp \left[ -\frac{\det {\bf \hat{\Gamma}}}{2N_0\hat{\Gamma}_{22}} 
\left( n_1-\bar{n}_1\right) ^2\right] \,\,.
\label{eq:Pproj}
\end{eqnarray}
We emphasize that the theoretical curves are produced with {\em no} fitting
parameters -- all quantities were computed from the model specifications
(i.e., $N_0$, $F$, and ${\bf M}$). For the ${\cal N}>2$ cases, it is difficult
to display full histograms. Instead, we only provide the comparison for the
averages and covariance matrices for three particular TSS's 
in tables~\ref{tab:I} and~\ref{tab:II}. 
As we see, the agreement is
excellent, well within the expected accuracy of the approximation, $O\left(
1/N_0\right) $, and the statistical errors, 
$O\left( 1/\sqrt{524,290}\right) $.

\section{Distributions containing dynamical information}
\label{sec:dyn}

In the previous section, we focused on the static distribution of the
populations in a TSS. In other words, we can compute (within the Gaussian
approximation) correlation functions that involve {\em any number} of
species, all ``at the same time.'' Here, we investigate the information
contained in the dynamics of the stochastic process, i.e., time-dependent
correlations. For an {\em evolving} population, a natural question is how
the system changes from one generation to the next. To probe this issue at a
quantitative level, let us consider two examples, the statistics of
``steps,'' 
\begin{equation}
\vec{s}\left( t\right) \equiv \vec{n}\left( t+1\right) -\vec{n}\left(
t\right) \;,
\label{eq:stepdef}
\end{equation}
and the correlation of these steps with the deviations from the average: 
$\vec{\varepsilon}\left( t\right) $.

In a stationary state, the mean-field prediction for the step size is
clearly zero ($\vec{s}^{\,{\rm MF}}\equiv 0$), as must also be the case for
the average $\left\langle \vec{s} \, \right\rangle ^{*}$. Nevertheless, we
expect a typical step size to be $O\left( \sqrt{N_0}\right) $. For a more
detailed picture, we may seek the step-size distribution (SSD) in the steady
state: ${\cal P}_s^{*}\left( \vec{s} \, \right) $. A precise definition is 
\begin{equation}
{\cal P}_s^{*}\left( \vec{s} \, \right) \equiv 
{\underset {t \rightarrow \infty } \lim }
\sum_{\vec{n}^{\prime },\vec{n}}\delta \left( \vec{s},\vec{n}^{\prime
}-\vec{n}\right) {\cal P}\left( \vec{n}^{\prime },t+1;\vec{n},t\right) \,\,,
\label{eq:PsDef}
\end{equation}
where ${\cal P}\left( \vec{n}^{\prime },t+1;\vec{n},t\right) $ is the 
{\em joint } probability for finding the system with population $\vec{n}$ 
at time $t$ {\em and} with $\vec{n}^{\prime }$ in the next step. From the 
master equation (\ref{eq:MEgen}), it is clear that this is just 
$R\left( \vec{n}^{\prime }|\vec{n}\right) {\cal P}\left( \vec{n},t\right) $, 
so that 
\begin{equation}
{\underset {t \rightarrow \infty } \lim }
{\cal P}\left( \vec{n}^{\prime },t+1;
\vec{n},t\right) =R\left( \vec{n}^{\prime }|\vec{n}\right) 
{\cal P}^{*}\left( \vec{n}\right) \,\,.
\end{equation}
Thus, once the steady-state distribution ${\cal P}^{*}\left( \vec{n}\right) $
is known, the SSD can be obtained from: 
\begin{equation}
{\cal P}_s^{*}\left( \vec{s} \, \right) =\sum_{\vec{n}}R\left(\vec{n}
+\vec{s}| \vec{n}\right) {\cal P}^{*}\left( \vec{n}\right) \,\,.  
\label{eq:Ps*}
\end{equation}

Applying this formalism to our case of coevolving species {\em without}
mutations, we can exploit all the approximations detailed in the previous
section, namely, a Gaussian for the stationary state 
(${\cal P}^{*}\left( \vec{n}\right) \cong {\cal P}_{\rm G}
\left( \vec{n}\right) $ 
of equation (\ref{eq:PG})), continuous variables for $\vec{n}$, and integrals 
instead of sums. Given equation (\ref{eq:P*}), the success of that scheme is 
implicitly dependent on the fact that 
$R\left( \vec{n}^{\prime }|\vec{n}\right) $ is also well approximated by 
a Gaussian 
\footnote{
In the same manner as for the stationary distributions, this property can be
derived from the definition of $R$ using straightforward, but tedious,
manipulations with the binomials. Here, we can treat this as an assumption,
the justification of which will be the agreement with simulation data.}. 
As a result, we need not carry out another lengthy analysis to conclude that 
${\cal P}_s^{*}\left( \vec{s}\right) $ should also be of the form 
\begin{equation}
{\cal P}_s^{*}\left( \vec{s} \, \right) \cong \left( 2\pi N_0\right) 
^{-{\cal N}/2}\left( \det {\bf \hat{\gamma}}\right) ^{1/2}\exp \left[ 
-\frac 1{2N_0}
s_I\hat{\gamma}^{IJ}s_J\right] \,\,.  
\label{eq:PsG}
\end{equation}
However, it is clear that the matrix ${\bf \hat{\gamma}}$ is distinct from 
${\bf \hat{\Gamma}}$, since the former must contain some ``dynamic''
information. (Note that the caret is to remind us that, despite the presence
of two superscripts, $\hat{\gamma}^{IJ}$ is of $O\left( 1\right) $.) Now,
within the context of simple Gaussians, ${\bf \hat{\gamma}}$ can be found by
computing its inverse 
\begin{equation}
{\bf \hat{g}\equiv \hat{\gamma}}^{-1} 
\end{equation}
which is just the second moment 
\begin{equation}
\left\langle s_Is_J\right\rangle _s\equiv \sum_{\vec{s}}s_Is_J{\cal P}%
_s^{*}\left( \vec{s}\right) \,\,.
\end{equation}
Let us first derive an exact formula for this quantity in terms of 
$\left\langle \bullet \right\rangle ^{*}$. Starting with 
\begin{eqnarray}
\left\langle s_Is_J\right\rangle _s &=&\sum_{\vec{s},\vec{n}}s_Is_JR\left( 
\vec{n}+\vec{s}|\vec{n}\right) {\cal P}^{*}\left( \vec{n}\right)
\label{eq:ss-def} \\
&=&\sum_{\vec{n}^{\prime },\vec{n}}\left( n_I^{\prime }-n_I\right) \left(
n_J^{\prime }-n_J\right) R\left( \vec{n}^{\prime }|\vec{n}\right) 
{\cal P}^{*}\left( \vec{n}\right) \,\,,
\end{eqnarray}
we use equations (\ref{eq:P*}) and (\ref{eq:fDef}) and arrive at the 
{\em exact} relation: 
\begin{equation}
\left\langle s_Is_J\right\rangle _s=2\left\langle n_In_J\right\rangle
^{*}-\left\langle f_In_J\right\rangle ^{*}-\left\langle n_If_J\right\rangle
^{*}\,\,.  \label{eq:ss-exact}
\end{equation}

\begin{figure}[t]
\includegraphics[angle=0,width=.47\textwidth]{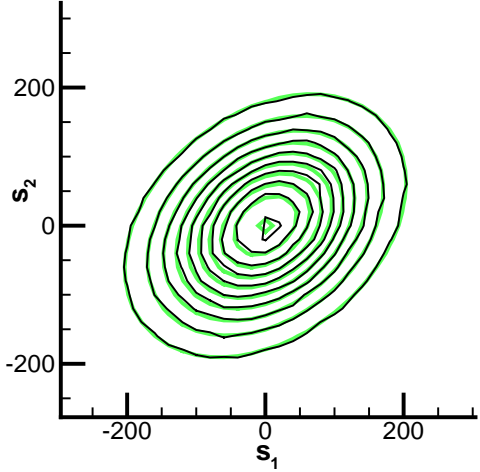}
\caption[]{
Contour plot of the 
theoretical Gaussian and simulated joint probability densities for the
steps, $s_1$ and $s_2$, equation (\protect\ref{eq:PsG}). 
The line types and parameters are the same as in
figure~\protect\ref{fig:pops}, 
and the simulated histogram was obtained from the same
simulation run as in that figure. 
The positive slope of the long axis indicates that the steps are
positively correlated.
}
\label{fig:diffs}
\end{figure}

Next, we exploit 
${\cal P}^{*}\left( \vec{n}\right) \cong {\cal P}_{\rm G}
\left( \vec{n}\right) $ 
for computing $\left\langle \bullet \right\rangle ^{*}$ and
apply equation (\ref{eq:Q*}). Thus, 
\begin{equation}
\left\langle n_In_J\right\rangle ^{*}\cong \bar{n}_I\bar{n}_J+N_0\hat{G}%
_{IJ} 
\label{eq:scov}
\end{equation}
and 
\begin{equation}
\left\langle f_In_J\right\rangle ^{*}\cong f_I\left(\bar{n}\right)\bar{n}_J
+N_0f_I^K\left( \bar{n}\right) \hat{G}_{KJ}
+ \frac{N_0}2f_I^{KM}\left( \bar{n}\right) \bar{n}_J\hat{G}_{KM}\,\,.  
\label{eq:mess}
\end{equation}
Thanks to equation (\ref{eq:nbar}), the first and the last terms in 
equation (\ref{eq:mess}) combine, so that 
\begin{equation}
\left\langle f_In_J\right\rangle ^{*}\cong \bar{n}_I\bar{n}_J
+ N_0f_I^K\left( \bar{n}\right) \hat{G}_{KJ}\,\,.  
\label{eq:<fn>}
\end{equation}
To the order kept here, $f_I^K\left( \bar{n}\right) $ is just 
$f_I^K\left( \vec{n}^{*}\right) =S_I^K$. So, collecting various items 
and using equation (\ref{eq:S}), we arrive at 
\begin{eqnarray}
\left\langle s_Is_J\right\rangle _s &\cong &N_0\left( 2\hat{G}_{IJ}
- S_I^K\hat{G}_{KJ}-\hat{G}_{IK}S_J^K\right) \\
&=&\,-N_0\left( \Lambda _I^K\hat{G}_{KJ}+\hat{G}_{IK}\Lambda _J^K\right) \,.
\end{eqnarray}
Finally, we relate this result to the Gaussian approximation 
equation (\ref{eq:PsG}) intended for the SSD, which provides 
\begin{equation}
\left\langle s_Is_J\right\rangle _s\cong N_0\hat{g}_{IJ}\,\,, 
\end{equation}
and obtain a simple equation for ${\bf \hat{g}}$: 
\begin{equation}
{\bf \hat{g}}=-\left( {\bf \Lambda \hat{G}+\hat{G}\Lambda }^T\right) \,\,.
\label{eq:g}
\end{equation}
As in the previous section, these predictions are well borne out in
simulations. For the same ${\cal N}=2$ case as above, 
we display a two-dimensional 
histogram of the step sizes in figure~\ref{fig:diffs}.
Note that this distribution is indeed quite different from
that for the populations (figure~\ref{fig:pops}). 
For the ${\cal N}>2$ cases, full
histograms are difficult to display and we only show the correlation
matrices ${\bf \hat{g}}$ and their Monte Carlo counterparts 
${\bf \hat{g}}_{\rm MC}$ in tables~\ref{tab:I} and~\ref{tab:II}.

Although the SSD probes the underlying dynamics, it does not contain {\em all} 
information of the stochastic (Ornstein-Uhlenbeck) process. In particular,
equation 
(\ref{eq:g}) shows that the antisymmetric part of ${\bf \Lambda \hat{G}}$
is ``missing.'' To remedy this shortcoming, we turn to another question
which naturally comes to mind: How are the steps ($\vec{s} \,$) correlated with
the deviations from the average ($\vec{\varepsilon}\equiv \vec{n}-\bar{n}$)?
Note that, since two quantities are involved ($\vec{s}$ and 
$\vec{\varepsilon}$), a full distribution associated with this question is 
slightly more involved than ${\cal P}^{*}\left( \vec{n}\right) $ or 
${\cal P}_s^{*}\left( \vec{s}\right) $. Nevertheless, within the 
approximation scheme we use, it would be a (generalized) Gaussian in the 
steady state. For simplicity, let us focus on, instead of the full 
distribution, 
only the normalized correlation of $\vec{s}$ with $\vec{\varepsilon}$: 
\begin{equation}
\hat{C}_{IJ}\equiv \left\langle s_I\varepsilon _J\right\rangle /N_0\,\,.
\end{equation}

By using this notation, we are displaying again our expectation that 
$\left\langle s_I\varepsilon _J\right\rangle $ is also of 
$O\left( N_0\right)$. Starting at the same point as equation (\ref{eq:ss-def}), 
we write 
\begin{eqnarray}
\left\langle s_I\varepsilon _J\right\rangle  &=&\sum_{\vec{s},\vec{n}}
s_I\varepsilon _JR\left( \vec{n}+\vec{s}|\vec{n}\right) {\cal P}^{*}\left( 
\vec{n}\right)   
\label{eq:se-def} \\
&=&\sum_{\vec{n}^{\prime },\vec{n}}\left( n_I^{\prime }-n_I\right) \left(
n_J-\bar{n}_J\right) R\left( \vec{n}^{\prime }|\vec{n}\right) {\cal P}^{*}
\left( \vec{n}\right) \,\,.
\end{eqnarray}
Since 
$\left\langle \vec{n}^{\prime }\right\rangle ^{*}
=\left\langle \vec{n}\right\rangle ^{*}$, this reduces to 
\begin{equation}
\left\langle s_I\varepsilon _J\right\rangle =\left\langle
f_In_J\right\rangle ^{*}-\left\langle n_In_J\right\rangle ^{*}\,\,,
\label{eq:se-exact}
\end{equation}
which is again an {\em exact} relationship. Repeating the analysis in the
previous paragraph, we find, in the Gaussian approximation, a simple result: 
$\left\langle s_I\varepsilon _J\right\rangle =\,N_0\Lambda _I^K\hat{G}_{KJ}$%
, or 
\begin{equation}
\,{\bf \hat{C}=\Lambda \hat{G}}\,.  \label{eq:se}
\end{equation}
Unlike the step-size covariance, this matrix is not symmetric in general and
contains the full information of the dynamics (to this order of our
approximation). Note also that, since ${\bf \Lambda }$ is negative and 
${\bf \hat{G}}$ is positive, this correlation is typically negative. This 
simply reflects a restoring dynamics: steps and deviations from the mean 
tend to be opposite. In tables~\ref{tab:I} and~\ref{tab:II}, 
we see that there is excellent agreement 
between the predicted ${\bf \hat{C}}$'s and their counterparts from 
simulations.

Another commonly studied correlation is the connected two-time function,
$\left( \langle n_I(t+\tau) n_J(t) \rangle 
- \langle n_I \rangle \langle n_J \rangle \right) / N_0$, 
which, in the stationary state, is just 
\begin{equation}
\hat{\Xi}_{IJ}(\tau)  
= \langle \varepsilon_I(t+\tau) \varepsilon_I(t) \rangle  / N_0
\;.
\label{eq:xi1}
\end{equation}
Within the Gaussian approximation, equation (\ref{eq:Langevin}) 
can be used repeatedly to 
express $\varepsilon (t+\tau)$ in terms of $\varepsilon(t)$ and the noise, 
$\eta_I(t+t')$ at each time step. 
Since $\langle \eta_I(t+t') \varepsilon(t) \rangle 
= \langle \eta_I(t+t') \rangle \langle \varepsilon(t) \rangle = 0$ 
for all $t'$, we find 
\begin{equation}
{\bf \hat{\Xi}}(\tau) = {\bf {S}}^\tau {\bf \hat{G}}
\;.
\label{eq:xi2}
\end{equation}

\section{Concluding remarks}

\label{sec:Concl}

Since we have presented a considerable amount of mathematical details above,
it is worthwhile to provide a short summary. In the model presented in 
\cite {RIKZIA03}, the entire evolution depends only on the parameters, 
$\mu$, $F$ , $N_0$ , and $M_{IJ}$.  
Strictly speaking, the only stationary state corresponds to total extinction 
($\vec{n}=0$). But, if mutation is suppressed ($\mu =0$), then, not only does
the configuration space break up into many sectors, but there will also be
non-trivial long-lived ($O\left( \rme^{N_0}\right) $ generations) states.
Referred to as ``truly stationary states,'' these communities consist of a
fixed number (${\cal N}$ $\geq 1$) of species. Typically, the population of
each species is $O\left( N_0\right) $. Provided we are not near a ``critical
point,'' the full distribution of these populations can be studied by a
systematic approximation scheme, starting with a multivariate Gaussian, 
parametrized by its mean and covariance matrix. For each TSS, our theory
predicts these parameters and thus, the full distribution. Specifically,
from the set of parameters, 
$\mu$, $F$ , $N_0$ , and $M_{IJ}$,  
we can compute $\vec{n}^{*}$, 
${\bf \hat{H}}$, and ${\bf S=1-\Lambda }$ 
(equations \ref{eq:fix3}-\ref{eq:A2}).
Then, at the lowest order in our approximation scheme, the mean $\bar{n}$
and the covariance matrix ${\bf \hat{G}}$ are given explicitly 
(equations \ref{eq:nbar=n*} and \ref{eq:GSH}-\ref{eq:explicitG}). 
In addition to 
this ``static'' aspect of the steady state, we also presented two ways of
characterizing the ``dynamic'' aspect. One is the full distribution of sizes
of steps (changes in the populations in a single time step, denoted by 
$\vec{s} \, $). Well approximated by a Gaussian also, this distribution 
has zero mean and covariance ${\bf \hat{g}}$ , which is given explicitly by 
equation (\ref{eq:g}). The other is the correlation of the steps with the 
populations just before the step (specified by the deviations from the 
average, $\vec{\varepsilon} \, $). Denoted by ${\bf \hat{C}}$, this correlation 
is given explicitly by equation (\ref{eq:se}). Finally, we have shown that 
there is excellent 
agreement between these analytical predictions and Monte Carlo simulations 
in ten typical TSS communities (data for three of which are shown in 
tables~\ref{tab:I} and~\ref{tab:II} 
and figures~\ref{fig:pops} and~\ref{fig:diffs}).

With a solid understanding of the steady-state properties of stable
communities in the absence of mutations, our next goal is a study of
quasi-steady states of populations with low mutation rates. To carry out
a serious analysis with $\mu >0$, we must define the restrictions on $R$
carefully. Otherwise, it would be impossible to find a satisfactory solution
to the equation 
\begin{equation}
{\cal P}^{*}\left( \vec{n}^{\prime }\right) = \sum 
R\left( \vec{n}^{\prime }|\vec{n}\right) 
{\cal P}^{*}\left( \vec{n}\right) \;. 
\end{equation}
As in the case for $\mu =0$, we may seek an approximate solution, in the
form of a product of a Gaussian distribution for the dominant species and
exponential distributions for the minority mutants \cite{ZIA04}. Provided
this program is successful, the next step would be to study the probability
of this type of QSS having ``dangerous'' mutants at the higher order of 
$\mu$, which hopefully will lead to some understanding of the distribution of
QSS lifetimes \cite{Krug03} 
and the presence of $1/f$ noise in power spectral densities.
Beyond this step, perhaps sophisticated renormalization-group techniques can
be marshalled to account for the self-similar patterns displayed, as well as
to identify universality classes for such behaviour. Needless to say, even
for such a simple model of coevolution, much remains to be explored.

\ack

We acknowledge enlightening discussions with many colleagues, in particular,
HW Diehl, BU Felderhof, K~Jain, J~Krug, AJ McKane, and L~Sch\"{a}fer. 
One of us (RKPZ) thanks HW Diehl for his hospitality at the
University of Duisburg-Essen (Germany) where some of this work was carried
out. This research is supported in part by the Alexander von Humboldt
Foundation (Germany), Florida State University through the Center for
Materials Research and Technology and the School of Computational Science
and Information Technology, and grants from the US National Science
Foundation (DMR-0088451, DMR-0120310, and DMR-0240078).

\appendix
\section{Mathematical Detail}

Starting from the general expression for a discrete-time Markov process, 
\begin{equation}
{\cal P}\left( \vec{n}^{\prime },t+1\right) =\sum_{\vec{n}}
R\left( \vec{n}^{\prime }|\vec{n}\right) 
{\cal P}\left( \vec{n},t\right) \;,  
\label{eq:A1}
\end{equation}
we write the mean-field approximation for the evolution equation as 
\begin{equation}
\vec{n}^{{\rm MF}}\left( t+1\right) =
\vec{f}\left( \vec{n}^{{\rm MF}}\left(t\right) \right) \;,  
\label{eq:A2'}
\end{equation}
where the {\em functional form} of $\vec{f}$ is given by 
\begin{equation}
\vec{f}\left( \vec{n}\right) =\sum_{\vec{n}^{\prime }}\vec{n}^{\prime}
R\left( \vec{n}^{\prime }|\vec{n}\right) \;.  
\label{eq:A3}
\end{equation}
Note that the components of $\vec{n}$ play the role of coordinates in equation 
(\ref{eq:A1}) and
take only non-negative integer values in models of population dynamics.
However, there is no guarantee that $\vec{f}$ will be an integer in 
equation (\ref{eq:A2'}), 
so that we must allow $\vec{n}^{{\rm MF}}$ to be 
continuous variables (functions of $t$). As we will see below, our analysis 
is much simplified if we also assume $\vec{n}$ to be continuous.

Focusing only on simple fixed points (as opposed to fixed cycles involving
two or more points) of the mean-field theory, we denote a FP by $\vec{n}^{*}$. 
(To keep the notation from being too cumbersome, we suppress the superscript 
$^{{\rm MF}}$ and only keep in mind that $\vec{n}^{*}$ represents just 
${\cal N}$ real numbers.) The equation for determining $\vec{n}^{*}$ is 
\begin{equation}
\vec{n}^{*}=\vec{f}\left( \vec{n}^{*}\right) \,\,.  
\label{eq:n*}
\end{equation}
Anticipating the need for the associated stability matrix, we write it as 
\begin{equation}
{\bf S\equiv }\left. \vec{\nabla}\vec{f} \, \right| _{\vec{n}^{*}}\,\,,
\label{eq:Sdef}
\end{equation}
where $\vec{\nabla}$ stands for derivatives with respect to $\vec{n}$ .
Explicitly, the matrix elements of ${\bf S}$ are 
\begin{equation}
S_I^J=\left. \partial f_I/\partial n_J\right| _{\vec{n}^{*}}  \label{eq:S'}
\;,
\end{equation}
clearly asymmetric in general. As in our earlier study, we restrict
ourselves to stable fixed points, so that the real parts of the eigenvalues
of ${\bf S}$ (denoted by ${\bf 1}+{\bf \hat{\Lambda}}$ in \cite{RIKZIA03})
should lie in the interval $(-1,1)$. Indeed, we restrict our attention
here to isolated FP's far from others, so that no (real part of any)
eigenvalue is close to unity.

Turning to the full stochastic problem, we consider a stationary
distribution ${\cal P}^{*}\left( \vec{n}\right) $, which satisfies 
\begin{equation}
{\cal P}^{*}\left( \vec{n}^{\prime }\right) =
\sum_{\vec{n}} R\left( \vec{n}^{\prime }|\vec{n}\right) 
{\cal P}^{*}\left( \vec{n}\right) \,\,.
\label{eq:P*ap}
\end{equation}
Therefore, it is a right eigenvector of the matrix 
$R\left( \vec{n}^{\prime}|\vec{n}\right) $ with unit eigenvalue. Its 
existence is guaranteed (though not necessarily its uniqueness, in general) 
by the presence of a left eigenvector of unit eigenvalue 
(i.e., $u\left( \vec{n}\right) =1 \,\, \forall \vec{n}$), 
thanks to having probability conserving rates. If ${\cal P}^{*}$
is known, then we can compute the stationary averages of all quantities via
equation (\ref{eq:<dot>}). A different approach is to study the equations
satisfied by these averages. Multiplying equation (\ref{eq:P*ap}) by a quantity
and summing, we find, in general, no closed equation. Instead, a tangle of
infinitely many coupled equations appear. Examples are 
\begin{equation}
\left\langle n_I\right\rangle ^{*}
=\left\langle f_I\left( \vec{n}\right)\right\rangle ^{*}
\quad \mathrm{and}\quad 
\left\langle n_In_J\right\rangle^{*}
=\left\langle H_{IJ}\left( \vec{n}\right) \right\rangle ^{*}\,\,,
\label{eq:f+H}
\end{equation}
where $f_I$ is an element of $\vec{f}$, given by equation (\ref{eq:A3}), and 
\begin{equation}
H_{IJ}\left( \vec{n}\right) \equiv \sum_{\vec{n}^{\prime }}
n_I^{\prime}n_J^{\prime }R\left( \vec{n}^{\prime }|\vec{n}\right) \,\,.  
\label{eq:Hdef}
\end{equation}
In both cases, the right-hand sides contain expectation values of all powers of 
$\vec{n}$, so that infinitely many equations are needed to close the set.

In practice, for generic situations such as those described in the main
text, we may approximate ${\cal P}^{*}$ by a Gaussian: 
\begin{equation}
\fl
{\cal P}^{*}\left( \vec{n}\right) \cong {\cal P}_{\rm G}
\left( \vec{n}\right)
=\left( 2\pi N_0\right)^{-{\cal N}/2}
\left( \det {\bf \hat{\Gamma}}\right)^{1/2}
\exp \left[ -\frac 1{2N_0}\left( n_I-\bar{n}_I\right) \hat{\Gamma}^{IJ}
\left( n_J-\bar{n}_J\right) \right] \,\,,  
\label{eq:PGap}
\end{equation}
where the unknowns $\bar{n}_I$ and $\hat{\Gamma}^{IJ}$ are to be determined.
Naturally, we expect that the center of the 
Gaussian lies close to the mean-field FP, 
$\vec{n}^{*}$. This scenario is corroborated by our case studies
above and our analysis below. Indeed, this approximation can be used as the
starting point of a systematic expansion, which relies on having a large
parameter ($N_0$) controlling the fluctuations to 
$O\left( \sqrt{N_0}\right)$. In this context, $\bar{n}$ and 
${\bf \hat{\Gamma}}$ are assumed to be $O\left( N_0\right) $ and 
$O\left( 1\right) $, respectively. In addition to playing the role of 
ordering a systematic expansion, a large $N_0$ (and so, large $\vec{n}^{*}$) 
provides two further simplifications: (i) $\vec{n}$ may be promoted to 
continuous variables so that sums can be replaced by integrals, and 
(ii) integration can be extended to $\left[ -\infty ,\infty\right] $ 
while only incurring errors of $O\left( \rme^{-\sqrt{N_0}}\right) $.

Once $\bar{n}_I$ and $\hat{\Gamma}^{IJ}$ are determined, then we can set up
expansions for the averages of any quantity $Q$: 
\begin{equation}
\left\langle Q(\vec{n})\right\rangle ^{*}\equiv 
\sum_{\vec{n}} Q(\vec{n}){\cal P}^{*}\left( \vec{n}\right) \,\,,  
\label{eq:Q1}
\end{equation}
starting with 
\begin{equation}
\int Q(\vec{n}){\cal P}_{\rm G}
\left( \vec{n}\right) {\rm d} \vec{n} \,\,.  
\label{eq:Q2}
\end{equation}
Given that $n_I$ does not deviate far from $\bar{n}_I$, such integrals can
be handled by first expanding $Q$ around $\bar{n}_I$: 
\begin{equation}
Q(\vec{n}) = \bar{Q} + 
Q^I\left( n_I-\bar{n}_I\right) +
\frac{1}{2} Q^{IJ}\left( n_I-\bar{n}_I\right) \left( n_J-\bar{n}_J\right) 
+ \ldots \;,  
\label{eq:Qexpand}
\end{equation}
where the summation convention is used and 
\begin{eqnarray}
\bar{Q} &\equiv &\left. Q\left(\vec{n}\right) \right| _{\vec{n}=\bar{n}}\;,
\label{eq:Q3a} \\
Q^I &\equiv &\left. \frac{\partial Q\left( \vec{n}\right) }{\partial n_I}
\right| _{\vec{n}=\bar{n}} \;,  
\label{eq:Q3b} \\
Q^{IJ} &\equiv &\left. \frac{\partial ^2 Q\left( \vec{n}\right) }
{\partial n_I\partial n_J}\right| _{\vec{n}=\bar{n}} \;.  
\label{eq:Q3c}
\end{eqnarray}
The integral can now be performed, so that 
\begin{equation}
\left\langle Q(\vec{n})\right\rangle ^{*}\cong \int Q(\vec{n})
{\cal P}_{\rm G}\left( \vec{n}\right) {\rm d} \vec{n}
=\bar{Q}\,+\frac{N_0}2Q^{IJ}\hat{G}_{IJ}+\ldots \;,
\label{eq:Q*}
\end{equation}
where ${\bf \hat{G}}={\bf \hat{\Gamma}}^{-1}$. There is no need to be
alarmed at the factor $N_0$ in the second term. Since $Q^{IJ}$ involves two
derivatives with respect to $\vec{n}$, 
it is typically of the order of $1/N_0^2$
compared to $\bar{Q}$. Thus, the second term in this expansion is 
$O\left(1/N_0\right) $ compared to the first.

With this machinery, we seek equations to fix $\bar{n}_I$ and ${\bf \hat{G}}$. 
From equation (\ref{eq:n+C}), we have, with $Q=f_I$, 
\begin{equation}
\bar{n}_I=\left\langle f_I\left( \vec{n}\right) \right\rangle ^{*}\cong
f_I\left( \bar{n}\right) +\frac{N_0}2f_I{}^{JK}\hat{G}_{JK}+\ldots \,\,.
\label{eq:nbar}
\end{equation}
Again, we expect the second term to be $O\left( 1/N_0\right) $ compared to
the first, so that we may write an expansion for $\bar{n}_I$: 
\begin{equation}
\bar{n}_I=\bar{n}_I^{(0)}+\bar{n}_I^{(1)}+\ldots \,\,.  
\label{eq:n-exp}
\end{equation}
Inserting this back into equation 
(\ref{eq:nbar}) and noting equation (\ref{eq:n*}), we find the expected 
\begin{equation}
\bar{n}_I^{(0)}=n_I^{*}\,\,.  
\label{eq:n0}
\end{equation}
Proceeding to the next order, we find 
\begin{equation}
\bar{n}_I^{(1)}=f_I^J\left( \bar{n}_I^{(0)}\right) \bar{n}_I^{(1)}
+\frac{N_0}{2} f_I^{JK}\hat{G}_{JK}\,\,,  
\label{eq:n1G}
\end{equation}
where all functions of $\vec{n}$ in the last term can be evaluated at 
$\bar{n}_I^{(0)}$. To save notation, we will just write $\hat{G}_{JK}$ 
in lieu of the more explicit form: $\hat{G}_{JK}^{(0)}$. We recognize that 
$f_I^J\left(\bar{n}_I^{(0)}\right) $ is just the stability matrix $S_I^J$ 
and write 
\begin{equation}
\bar{n}_I^{(1)}=\frac{N_0}2V_I^Mf_M{}^{JK}\hat{G}_{JK} \;, 
\label{eq:n-G}
\end{equation}
where 
\begin{equation}
{\bf V\equiv }\left( {\bf 1-S}\right) ^{-1}  
\label{eq:Vdef}
\end{equation}
(the inverse of $-{\bf \tilde{\Lambda}}$ in \cite{RIKZIA03}). Let us 
emphasize that $V_I^M$ and $\hat{G}_{JK}$ are supposedly $O\left( 1\right) $, 
while $f_M{}^{JK}$ is $O\left( 1/N_0\right) $ so that the right-hand side of 
equation 
(\ref{eq:n-G}) is a quantity of $O\left( 1\right) $. This expression also
provides a precise meaning to the phrase ``no eigenvalue (of ${\bf S}$) is
close to unity,'' which appeared as a restriction on the FP's we study.

Extending this technique to $n_In_J$ , we consider 
\begin{equation}
N_0\hat{G}_{IJ}+\left\langle n_I\right\rangle ^{*}\left\langle
n_J\right\rangle ^{*}=\left\langle n_In_J\right\rangle ^{*}=\left\langle
H_{IJ}\left( \vec{n}\right) \right\rangle ^{*}\,\,,  
\label{eq:nn-exact}
\end{equation}
from equations (\ref{eq:n+C}) and (\ref{eq:f+H}). Again, using the Gaussian
approximation for both sides, we arrive at 
\begin{equation}
N_0\hat{G}_{IJ}+\bar{n}_I\bar{n}_J\cong H_{IJ}\left( \bar{n}\right) 
+\frac{N_0}2H_{IJ}{}^{KM}\hat{G}_{KM}+\ldots \,\,.  
\label{eq:G-H}
\end{equation}
At the lowest order, $O\left( N_0^2\right) $, of this equation, internal
consistency will ensure that 
$H_{IJ}\left( \bar{n}^{(0)}\right) \cong \bar{n}_I^{(0)}\bar{n}_J^{(0)}$ 
and should provide a check for tedious, error-prone
computations. Furthermore, we can consider the {\em difference} : 
\begin{equation}
H_{IJ}\left( \vec{n}\right) -
f_I\left( \vec{n}\right) f_J\left( \vec{n}\right)  
\label{eq:h_IJ}
\end{equation}
and find that it is often one order lower (as explicitly shown in the main
text for our model). In other words, we have 
\begin{equation}
H_{IJ}\left( \vec{n}^{*}\right) -
f_I\left( \vec{n}^{*}\right) f_J\left( \vec{n}^{*}\right) 
\sim O\left( N_0\right) \,\,.
\end{equation}
To emphasize this property, we explicitly extract a factor $N_0$ and
define 
\begin{equation}
\hat{H}_{IJ}\left( \vec{n}\right) \equiv \left[ H_{IJ}\left( \vec{n}\right)
-f_I\left( \vec{n}\right) f_J\left( \vec{n}\right) \right] /N_0\,\,,
\label{eq:Htilde0}
\end{equation}
so that 
\begin{equation}
H_{IJ}\left( \bar{n}\right) 
=f_I\left( \bar{n}\right) f_J\left( \bar{n}\right) 
+N_0\hat{H}_{IJ}\left( \bar{n}\right) \,\,.  
\label{eq:Hff}
\end{equation}
Since we keep only terms of orders $N_0^2$ and $N_0$, we can evaluate 
$\hat{H}_{IJ}\left( \bar{n}\right) $ at $\vec{n}^{*}$ and define, 
for simplicity, 
\begin{equation}
\hat{H}_{IJ}\equiv \hat{H}_{IJ}\left( \vec{n}^{*}\right) \,\,.
\label{eq:Htilde}
\end{equation}
As a reminder, this quantity is of $O\left( 1\right) $. Next, to lowest
order needed (i.e., $O\left( 1\right) $), we find 
\begin{equation}
H_{IJ}{}^{KM}=S_I^KS_J^M+S_I^MS_J^K+f_If_J{}^{KM}+f_I{}^{KM}f_J{}\,\,,
\label{eq:H_IJKM}
\end{equation}
all evaluated at the FP. Inserting these expressions into 
equation (\ref{eq:G-H}), we obtain 
\begin{eqnarray}
\fl
N_0G_{IJ}+\bar{n}_I\bar{n}_J
&\cong& 
f_I\left( \bar{n}\right) f_J\left( \bar{n}\right) 
+ N_0\hat{H}_{IJ} \nonumber\\
&& +\frac{N_0}2 \left[S_I^KS_J^M+S_I^MS_J^K 
+ f_If_J{}^{KM}+f_I{}^{KM}f_J\right] G_{KM}+\ldots 
\,\,. 
\nonumber\\
\end{eqnarray}
Exploiting equation (\ref{eq:nbar}), we arrive at an equation for 
${\bf \hat{G}}$ : 
\begin{equation}
\hat{G}_{IJ}
=\hat{H}_{IJ}+\frac 12\left[ S_I^KS_J^M+S_I^MS_J^K\right] \hat{G}_{KM} \;.
\end{equation}
But, ${\bf G}$ is symmetric, so that the final equation, in matrix form,
reads 
\begin{equation}
{\bf \hat{G}-S\hat{G}S}^T={\bf \hat{H}\,\,.}  \label{eq:ansG}
\end{equation}
Here, ${\bf S}^T$ denotes the transpose of ${\bf S}$, and, from 
equation (\ref{eq:Hdef}), ${\bf \hat{H}}$ is necessarily symmetric. Again, let 
us emphasize that all quantities in this equation are $O\left( 1\right) $.
Unfortunately, there is no simple way 
to express ${\bf \hat{G}}$ 
in terms of the known matrices: ${\bf S}$ and ${\bf \hat{H}}$. 
A formal series can be written as 
\begin{equation}
{\bf \hat{G}}
=
{\bf \hat{H}+S\,\hat{H}} \thinspace {\bf S}^T+{\bf SS\,\hat{H}}
\thinspace {\bf S}^T{\bf S}^T \thinspace +...
\;,
\label{eq:formser}
\end{equation} 
equivalent to a recursion relation, 
\begin{equation}
{\bf \hat{G}}_n = {\bf \hat{H}} + {\bf {S}}{\bf \hat{G}}_{n-1}{\bf {S}}^T
\label{eq:recrel}
\end{equation}
with ${\bf \hat{G}}_0 = {\bf \hat{H}}$. 
It was this recursion method that was used to obtain numerical results for 
comparison with the Monte Carlo results in the figures and tables. 
Convergence to four significant digits was obtained with 
$n=20$ for ${\cal N} =2$, $n=30$ for ${\cal N} =3$, 
and $n=140$ for ${\cal N} =4$.

An explicit form for ${\bf \hat{G}}$ 
can be found when we examine this equation in the frame where 
${\bf S}$ is diagonal. To be explicit, take matrix elements of 
equation (\ref{eq:ansG}) with $u_a^I$, the left eigenvectors of ${\bf S}$, 
which satisfy 
\begin{equation}
u_a^IS_I^K=\lambda _au_a^I\qquad \mathrm{(no\;sum\;on\;}a) \;. 
\end{equation}
The result is 
\begin{equation}
\hat{G}_{ab}=\hat{H}_{ab}/\left( 1-\lambda _a\lambda _b\right) 
\qquad \mathrm{(no\;sum)} \;, 
\label{final}
\end{equation}
where 
\begin{equation}
\hat{G}_{ab}=\hat{G}_{IJ}u_a^Iu_b^J\qquad \mathrm{and}\qquad \hat{H}_{ab}
=\hat{H}_{IJ}u_a^Iu_b^J\,\,.  \label{eq:Hab}
\end{equation}
To find the original matrix elements, we apply the dual set $v_I^a$ (i.e.,
the right eigenvectors of ${\bf S}$, normalized by $v_I^au_b^I=\delta _b^a$)
and obtain 
\begin{equation}
\hat{G}_{IJ}=v_I^a v_J^b \hat{G}_{ab}
=\sum_{a,b}v_I^av_J^b\hat{H}_{ab}/\left(1-\lambda _a\lambda _b\right) \;.  
\label{eq:explicit}
\end{equation}

Finally, this expression again shows the importance of insisting that ``no
eigenvalue (of ${\bf S}$) is close to unity.'' In case one or more $\lambda $
approaches unity, the system would be labelled ``critical'' or
``multi-critical,'' in the language of phase transitions.
Away from such
points, the Gaussian approximation, along with its associated
Ornstein-Uhlenbeck process, is adequate. For further details, see, for
example, the books by van Kampen~\cite{vanKampen} and Risken~\cite{Risken}.

\end{document}